# Local shear waves attenuation measurements by the MR methods

T. Klinkosz[(a)], C.J. Lewa[(b)]
*Institute of Experimental Physics, University of Gdańsk, Wita Stwosza 57, Poland*
[(a)] e-mail: fiztom@univ.gda.pl, [(b)] tel./fax (048)(058)341-31-75, e-mail: fizcl@univ.gda.pl

**Abstract.** - The essential feature of the method is the employment of Elasto-Magnetic Resonance Spectroscopy (EMRS) for precise local measurements of the attenuation of shear elastic waves introduced into a biological sample. Such a measurement can be accomplished by combining the EMRS method with such methods, in which collective dislocations of spins are induced by external physical factors, such as variable electric field, strong magnetic field gradient or longitudinal elastic wave. Theoretical bases of the method presented, related with the external factors, are discussed in the present paper.

The imaging of the shear coefficient distribution (of longitudinal or shear waves) has become an important diagnostic tool and can be accomplished using the method of nuclear magnetic resonance EMRS [1,2], MRE – Magnetic Resonance Elastography [3], as well as ultrasonic methods (Sonoelasticity) [4]. In recent years, several new approaches were proposed for determining mechanical properties of tissues, as regards both magnetic resonance and ultrasonic methods [5]. Elastography has found numerous applications in medical diagnostics, e.g. for identification of neoplastic tumours of the breast, prostate, brain or liver, the rigidity of which generally differs markedly from that healthy surrounding tissues, in hypothermia, palpation of the brain, surgical interventions, sports medicine (e.g. for determination of muscular tension) etc.

The diagnostic potential and specificity of the methods mentioned above can be substantially expanded by determining additional physical parameters, such as viscosity or the coefficient of elastic waves attenuation in the matter. These parameters have so far been determined in several different ways: i) by finding the local shear waves length in the medium and, hence, evaluating the amplitude of the dislocation of the medium elements; as a result, by adjusting to a suitable theoretical model, elasticity coefficients of the medium for a shear wave could be determined, mainly by the MR methods [6], ii) by measuring the phase propagation velocity of shear waves and applying correlation techniques of calculations [7], or by ultrasonic methods based on the Doppler effect. Not long since, methods called "real-time elastography", employing 3-dimensional finite model of calculations [8], have been in use.

Changes in the shear waves attenuation coefficient provide valuable diagnostic information in cases of calcinosis or lesions caused by thermal tissue ablation. Heating of a biological tissue to high temperatures (above 65°C) results in significant changes in the attenuation coefficient. These results suggest that attenuation measurement may be a useful indicator of tissue damage during thermal therapy.

In the present report, a novel method has been proposed, which permits determination of the local value of the attenuation coefficient of a shear wave in the medium, based on the combination of the EMRS and external factors which induce collective dislocation of spins in the sample.

*Principles of the method.* - The method proposed is based on the measurement of phase changes, *DF*, of the transverse magnetization component. Transverse magnetization phase changes are proportional to the scalar product of the amplitude of the magnetic field gradient $\vec{G}_0$ and the effective displacement of paramagnetic elements, $\vec{z}_{eff}$, in the sample, caused by external factors shown in Table 1. The effective dislocation of spins is a superposition of the dislocation evoked by a shear wave propagating in the sample, $\vec{z}_x(t)$, and the dislocation induced by an external factor causing collective spin dislocation, $\vec{z}_F(t)$, within the whole sample.

The phase changes in the transverse magnetization component can be expressed as a superposition of changes in this phase resulting from spin dislocation caused by a shear elastic wave and those induced by an additional external factor:

$$\Delta\Phi = \Delta\Phi_x + \Delta\Phi_F \qquad (1)$$

When the magnetic field gradient is given by $\vec{G}(t) = \vec{G}_0 \sin(w_G t)$, spin dislocations are induced by the elastic wave $\vec{x}(t) = \vec{x}_0 e^{-mx} \sin(w_x t - kx + \varphi_x)$. Hence, the phase changes in the transverse magnetization component $\Delta\Phi_\xi$ after $N$ periods, $T_G$, of the magnetic field gradient $\vec{G}(t)$ are given by the relation [1]:

$$\Delta\Phi_x = \frac{\gamma N \vec{G}_0 \cdot \vec{z}_{x,0}}{w_G(a-1)} \exp(-m \cdot x) \cdot \cos\left[w_G(a-1)\left(t + \frac{NT_G}{2}\right) - kx + \varphi_x\right] \sin\left[w_G(a-1)\frac{NT_G}{2}\right] \qquad (2)$$

where: γ is the gyromagnetic ratio, $k$ is the wave number of the share wave, $t$ denotes time, $a = w_x/w_G$, $w_G$, $w_x$ are angular frequencies of the changes in the magnetic field gradient, $\vec{G}(t)$, and spin dislocations, $\vec{x}(t)$, induced by the shear wave, respectively; $\vec{z}_{x,0}$ is the amplitude of spin dislocations caused by the shear wave, $m$ is the attenuation coefficient of the shear wave, and $\varphi_x$ is the shear wave phase.
These phase changes described by eq. (1), displaying a resonance character (maximum changes for $w_G = w_x$) and spatial selectivity, are time-independent [1]. This facilitates precise determination of their spatial dependence, and allows to: i) calibrate the selected methods, ii) sweep the sample investigated by changing the phase of shear waves, iii) measure viscoelastic properties of the sample.

*Forms of the external factor.* - The external factor inducing the collective spin dislocation may originate, as already mentioned, from: i) an electric field [9], ii) a strong magnetic field gradient [10] (then, the Stern-Gerlach interaction is present), iii) a longitudinal elastic wave. Therefore, the following forms of this factor are possible. Expressions in Table 1 describe collective spin dislocations in the sample, $\vec{z}_i(t)$ (*i=E, SG, u*), induced by the respective factors. $\vec{E}_0$ denotes the amplitude of the electric field with frequency $w_E$ applied to the sample, $\vec{G}_{SG,0}$ is the amplitude of a strong sinusoidal magnetic field gradient with frequency $w_{SG}$, $\vec{z}_{u,0}(t)$ is the amplitude of a longitudinal elastic wave with frequency $w_u$, $\varphi_E$, $\varphi_{SG}$, $\varphi_u$ are phases of the electric field, magnetic field gradient, and the elastic wave, respectively, $\eta$ is the microviscosity of the medium, $a_{ef}$ is the effective radius of the displaced carriers of charge $q$, $m$ is the energy state of a chosen spin group, $\hbar = h/2\pi$, where $h$ is the Planck constant, $k_u$ is the wave number of the longitudinal waves, $\alpha_u$ denotes the attenuation coefficient of the longitudinal wave, and $b = w_E/w_G$, $k = w_u/w_G$, $T_{SG}$ is the period of the Stern-Gerlach strong magnetic field gradient. The forms of the factor forcing the collective movement of spins in the sample and corresponding expressions describing spin dislocations thus caused are given in Table 1.

TABLE 1. - *Types of the external factor inducing collective spin dislocations and expressions describing such dislocations in the sample.*

| No. | type of external factor | formulae for external factor | spin dislocation |
|---|---|---|---|
| 1 | electric field [3] | $\vec{E}(t) = \vec{E}_0 \sin(w_E t + j_E)$ | $\vec{z}_E(t) = \dfrac{q\vec{E}_0}{6pha_{ef}w_E} \sin(w_E t + j_E)$ |
| 2 | strong magnetic field gradient [10] | $\vec{G}_{SG}(t) = \vec{G}_{SG,0} \sin(w_{SG} t + j_{SG})$ | $\vec{z}_{SG}(t) = \dfrac{g\hbar m \vec{G}_{SG,0}}{6pha_{ef}w_G}(\cos j_{SG} - \cos(w_{SG}t + j_{SG}))$ |
| 3 | long elastic wave | $\vec{z}_u(t) = \vec{z}_{u,0} e^{-a_u y}(t)\sin(w_u t - k_u y + j_u)$ | $\vec{z}_u(t) = \vec{z}_{u,0} \sin(w_u t + j_u)$ * |

\* When the sample volume is much smaller than the wavelength of longitudinal waves. Then, a piston movement of the sample area investigated can be observed. For example, when the angular frequency, $\omega_u$, of the longitudinal waves amounts to $\omega_u = 2\pi \cdot 200$ Hz, the velocity, $c_u$, of these waves in a gel phantom is $c_u = 1500$ m/s and $\lambda_u = 7.7$ m. The dimension of the sample is L = 12 cm, i.e. L = $(1/64)\lambda_u$.

TABLE 2. - *Magnetization phase changes, $\Delta\Phi_F$, depending on the type of the external factor inducing collective spin dislocation in the sample.*

| nr | ext. factor | expression for $\Delta\Phi_F$ |
|---|---|---|
| 1 | $\vec{E}(t)$ | $\Delta\Phi_F = g\dfrac{qN\vec{G}_0 \cdot \vec{E}_0}{6pha_{ef}w_E w_G(b-1)}\cos(\tfrac{1}{2}w_G(b-1)(t+NT_G+1)+j_E)\cdot \sin(\tfrac{1}{2}w_G(b-1)(t+NT_G-1))$ |
| 2 | $\vec{G}_{SG}(t)$ | $\Delta\Phi_F = \dfrac{\hbar m}{3p^3 ha_{ef}}(gT_{SG}NG_{SG,0}\cos j_{SG})$ |
| 3 | $\vec{z}_u(t)$ | $\Delta\Phi_F = \dfrac{gN\vec{G}_0 \cdot \vec{z}_{u,0}}{w_G(k-1)}\cos\left[w_G(k-1)\left(t+\dfrac{NT_G}{2}\right)+j_u\right]\sin\left[w_G(k-1)\dfrac{NT_G}{2}\right]$ |

According to eq. (1), the observed resultant magnetization phase changes are a superposition of phase changes, $\Delta\Phi_x$, due to movement of spins forced by the shear wave and changes $\Delta\Phi_F$ caused by the external factor described by respective expressions in Table 2 (third column).

Figure 1 illustrates phase changes induced by a low-frequency shear wave and, additionally, by a variable electric field in which the sample is placed. Assuming that the electric field, $E(t)$, is parallel to the magnetic field gradient, $G(t)$, the following form of the phase change of the transverse magnetization component is obtained:

$$\Delta\Phi = \dfrac{gN\vec{G}_0 \cdot \vec{z}_{x,0}}{w_G(a-1)}\exp(-m\cdot x)\cdot\cos\left[w_G(a-1)\left(t+\dfrac{NT_G}{2}\right)-kx+j_x\right]\sin\left[w_G(a-1)\dfrac{NT_G}{2}\right]+ \\ +g\dfrac{qN\vec{G}_0 \cdot \vec{E}_0}{6pha_{ef}w_E w_G(b-1)}\cos(\tfrac{1}{2}w_G(b-1)(t+NT_G+1)+j_E)\cdot\sin(\tfrac{1}{2}w_G(b-1)(t+NT_G-1)) \quad (3)$$

The phase changes described by eq. (3) also display the resonance character. Here, it is a triple resonance – the maximum changes for $w_G = w_E = w_x$ and spatial selectivity are time-independent, thus facilitating precise determination of their spatial dependence, and allowing to: i) calibrate the selected methods, ii) sweep the sample investigated by changing the phase or frequency of factors which force the movement, iii) measure viscoelastic properties of the sample, iv) image the spatial distribution of the attenuation coefficient of the shear wave with angular frequency $w_x$ propagated in the sample; when the distribution of elasticity and electrical mobility of the charged paramagnetic elements (ions, chemical radicals) are known. The phase changes of the transverse magnetization component versus valuse of *a* and *b* are presented in Fig. 2.

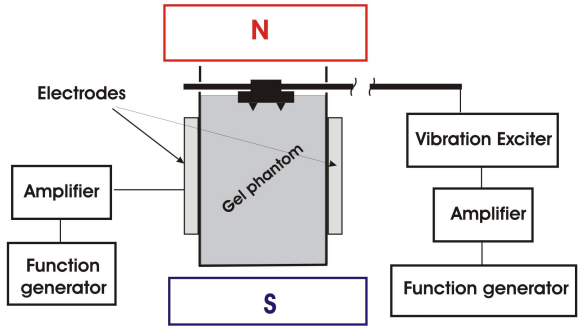
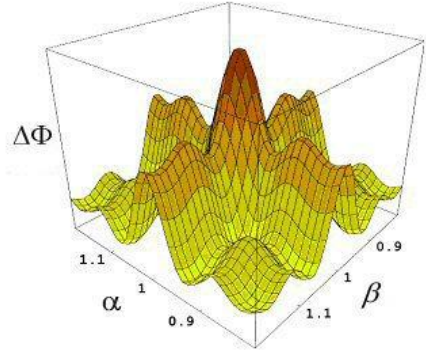

Fig. 1. Fig. 2.

Fig. 1. - Experimental setup for the combination of EMRS and EMMRS.

Fig. 2. - 3D plot of *DF* (3) as a function of relative frequencies: $a = w_x/w_G$, $b = w_E/w_G$. Maximum changes in transverse magnetization occur when $a = b = 1$, i.e. $w_G = w_E = w_x$.

*Measurements of attenuation coefficient* **m** - Since the external force, $\vec{F}(t)$, causes the collective movement of spins in the whole sample volume, i.e. irrespective of the position on the shear wave propagation direction (the *x*-axis), the measurement can be carried out as follows. Let us choose a local minimum (or local maximum) of the function *DF* as its characteristic point (see Fig. 3). *Dx* is the spatial resolution of the method (the minimum segment of the sample, along which the phase changes). Magnetization phase changes, $dF = DF_1 - DF_2$, can be measured by changing the elastic wave phase $Dj_x = j_2 - j_1$. Phases $j_1$, $j_2$ are different for individual runs of functions $DF_1(x)$, $DF_2(x)$. One should note the proportionality of the *dF* changes to the external factor, $\vec{F}(t)$, inducing the collective spin movement within the sample. We have succeeded to dislocate the chosen minimum of the function (i.e. we dislocated the point which was the source of the MR response) by *Dx*, which is not the spatial resolution of images obtained (it is not a pixel size). In practice, it is convenient to change the phase $j_x$ by the delay in triggering the imaging sequence of the MR tomograph. Next, by changing the amplitude of the factor causing dislocation of spins within the whole sample volume, $\vec{z}_{F,0}$, which is performed by changing the value of $\vec{F}_0$, the chosen minimum can be reduced to zero value. Since the changes in the transverse magnetization component are proportional to the dislocation induced by the external factor, $\vec{F}(t)$, the magnetization phase changes can be measured by measuring the amplitude, $\vec{F}_0$, or by the direct measurement of magnetization phase changes, *DF*. By repeating this procedure many times, one can sweep the selected area of the sample and determine its viscoelastic properties and attenuation coefficient of a shear wave.

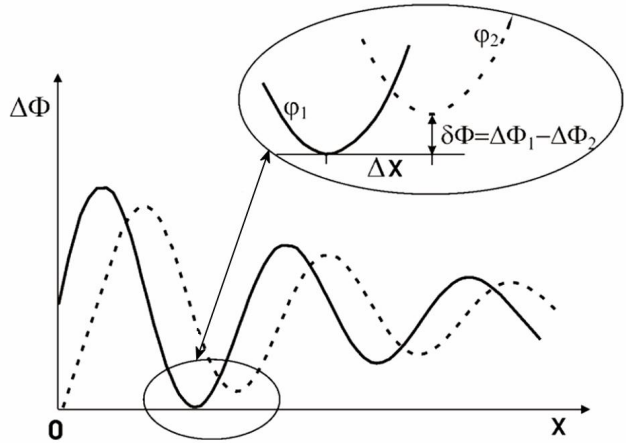

Fig. 3. - The method of measuring the shear wave attenuation coefficient, $m$, in the sample

*Expected results.* - Figure 4 shows the experimental setup designed to measure the maximum amplitude of vibrations for the plate inducing the shear wave, depending on the frequency of vibrations. The amplitude was determined by means of a miniature Bruel & Kjaer, Type 4393 accelerometer. The results are summarized in Table 3 (second column).

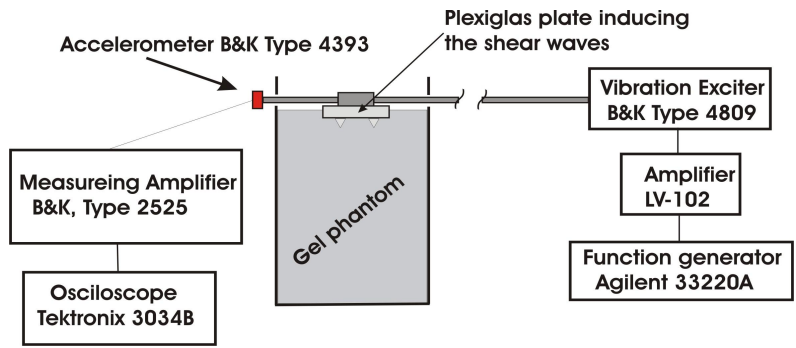

Fig. 4. - Experimental setup for determining the maximum amplitude of the shear wave for different frequencies of mechanicel vibrations.

The MR images (Fig. 5) show a shear wave propagating in the gel phantom. As seen, the wave penetrates the sample to a depth from several to twenty centimeters.

In order to analyze the potential of the method, being a combination of the MR elastography with external factors inducing the collective spin dislocation in the sample, let us reduce expression (1) describing the changes in the transverse magnetization component. Phase changes occurring at the boundary between the visco-elastic medium and the plate inducing the shear wave are shown in Fig. 4. In this area, maximum amplitude, the greatest spin dislocation can be observed. With such an approach to the problem, the exponential factor related with the wave attenuation in the medium does not have to be taken into account in expression (2). Hence, the magnetization phase change, *DF*, for protons, expressed in degrees, can be given in the gollowing form:

$$\Delta\Phi[\text{deg}] \approx \frac{1}{2} g N T_G (\vec{G}_0 \cdot \vec{z}_{x,0}) \cdot \frac{180^\circ}{p} \approx 21.28 \cdot 10^6 \cdot N T_G (\vec{G}_0 \cdot \vec{z}_{x,0}) \cdot \frac{180^\circ}{p} \quad (4)$$

For example, let us consider the amplitudes for three frequencies: $n_{SW,1}=100\ Hy$, $n_{SW,2}=250\ Hz$, $n_{SW,3}=500Hz$. Assuming in addition $N=4$ – four periods of the sinusoidal magnetic field gradient with the amplitude
$G_0=12\ mT/m$ and frequency equal to the frequency of the shear wave in the MR imaging sequence, the following values for changes in the magnetization phase, $\Delta\Phi$, in degrees are obtained (Table 3, third column).

TABLE 3. - *Maximum value of the spin displacement amplitude and corresponding magnetization phase changes calculated according to equation (4) and expressed in degrees.*

| Frequency of the shear wave $\nu_{SW}$ [Hz] | Maximum amplitude observed $\zeta_{SW}$ [mm]±0.01mm | Changes in the magnetization phase $\Delta\Phi_\xi$ [°] |
| --- | --- | --- |
| 100 | 0.27 | 158,2 |
| 250 | 0.22 | 51.5 |
| 500 | 0.03 | 3.5 |

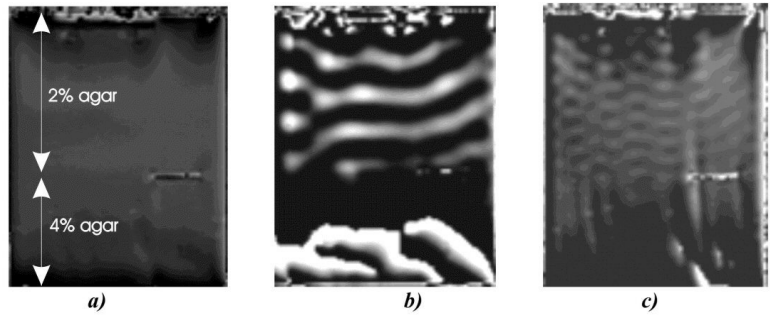

Fig. 5. - Phase images of the shear waves with frequencies: *a)* 0 Hz - without shear waves Hz, *b)* 250 Hz, *c)* 500 Hz propagating in 2% agar fantom (from the inducing plexiglass plate to 8 cm deep) and 4% agar-gel phantom (from 8 cm to 15 cm deep). The dimensions of the phantom: 12x12x15 cm. It can be seen that the penetration depth of the shear waves is about 7 cm or even more. *(Autors of images: G. Madelin, E. Thiaudiere, T. Klinkosz - June 2004 Magnetic Resonance Center, CNRS – University Victor Segalen Bordeaux 2)*

Assuming that the shear wave attenuation coefficient for agar gel amounts to about 50Np/m [7], the amplitude of spin dislocation is of the order of 17.3 µm for the frequency 500 Hz at a depth of 7 cm from the surface of the plate. This means that theoretically, according to eq. (1), we are able to observe phase changes $DF \approx 0.1$ degree. Using a tomograph with higher values of the magnetic field gradient one can expect to observe markedly lower amplitudes of vibrations. Dislocations of the order of $Dz_{x0} \approx 50$ nm at an accumulation of $N=10$ were observed at the magnetic field gradient $G_0 \approx 23mT/m$ and shear wave frequency $n_{SW} \approx 500$ Hz [11]. According to relation (4), this corresponds to the magnetization phase change $DF \approx 0.015$ degree. Thus, it can be inferred that in the case considered it should be possible to determine small changes in the attenuation coefficient for the shear wave of the order of $4 \cdot 10^{-2}$ dB/cm. On the other hand, the application of one of the external factors will improve the spatial resolution in comparison to the MRE method and allow determination of the local coefficient, i.e. the attenuation along a distance which is a fraction of the shear wave length.

*Conclusions.* - This method, based on the combination of EMRS and other physical factors which induce collective spins displacement (EMMRS in this case), allows precise determination and imaging of the attenuation coefficient by the dislocation-phase transformation of the transverse magnetization component, *DF*, thus substantially improving the research and application potential of the MRI methods. The elastic wave phase change causes spatial dislocation of the observation point (Fig. 3) which is the source of the MR response and, hence, it makes possible to measure the local elastic wave attenuation coefficient in the medium. The limitations involved in the application of the method result from the dependence of the magnetization phase changes on the scalar product ($\vec{G}_0 \cdot \vec{z}_{eff}$). In view of the attenuation of the shear elastic wave in the sample, the spatial resolution of the method will decrease with decreasing amplitude $\vec{z}_x(t)$.

*Acknowledgements.* - This research was supported by Polish Committee for Scientific Research, contract No. 2P03B06324.